\documentclass{pasj00}
\begin{document}
\SetRunningHead{Cao, G and Wang, J. C}{Particle Acceleration And Emission Processes In Mrk 421}
\title{Particle Acceleration And Emission Processes In Mrk 421}
\author{Gang \textsc{Cao}}
\affil{National Astronomical Observatories, Yunnan Observatory,
Chinese Academy of Sciences,  Kunming 650011, China} \email{gcao@ynao.ac.cn}
\author{Jiancheng \textsc{Wang}}
\affil{National Astronomical Observatories, Yunnan Observatory,
Chinese Academy of Sciences,  Kunming 650011,
China} \email{jcwang@ynao.ac.cn}
 %Do NOT move this preamble from here!

\maketitle
\begin{abstract}
We investigate the X-ray and $\gamma$-ray flares of Mrk 421 on 2008 June 6-15 using the synchrotron self-Compton(SSC) model with electron acceleration, in which an evident correlation between the  X-ray and $\gamma$-ray bands appears, while no significant correlation between the optical and X-ray band is observed. We argue that the emission from Mrk 421 may originate from two different components. One is the steady component from the outer region that is mainly attributed to the optical band, in which the electrons are accelerated by first-order Fermi acceleration mechanism. We use a steady electron spectrum to produce the synchrotron self-Compton emission. The other is the variable component from the inner region, in which the electrons are accelerated by the stochastic acceleration process. We use the time-dependent SSC model to produce the emission from the variable component. We suggest that the flares are due to the hardening of the electron spectrum under the process of the stochastic acceleration, which leads to the hardening of the observed spectrum in the X-ray and $\gamma$-ray bands. Furthermore, we find that the energy densities of electrons and magnetic fields are near equipartition in both jet regions.
\end{abstract}

\section{Introduction}
Blazars are the most extreme subclass of Active Galactic Nuclei (AGNs). Their radiation are thought to originate in a relativistic jet oriented at a small angle with respect to the line of the sight. They usually show strong and rapid variability, high and variable polarization, and super-luminal motions, etc. The spectral energy distributions (SEDs) of blazars are dominated by non-thermal emission and consist of two distinct, broad components. It is generally agreed that the low component of their SEDs is produced by synchrotron emission from relativistic electrons, while the origin of high energy component is still a matter of debate.  There are two classes of models to explain high energy emission: leptonic model and hadronic model. In the leptonic model, the high energy emission is produced by inverse Compton (IC) scattering of the soft photon fields (e.g., B\"{o}ttcher 2007), in which soft photons are the synchrotron photons within the jet (the SSC process; Maraschi et al. 1992; Bloom and Marscher 1996 ) or the photons external to jet (the EC process). These external photons are the UV accretion disk photons \citep{der93} or the accretion disk photons reprocessed by broad line region clouds \citep{sik94} or the infrared photons from the dust torus \citep{bla00}. In the hadronic model, the high energy emission originates from proton synchrotron or photon-hadronic interaction \citep{man93,muc00,bot09}. A way to distinguish the different emission models is to study variability of blazars. The one-zone SSC model predicts a significant X-ray/$\gamma$-ray correlation. However, the hadronic model is challenged by the observed X-ray/$\gamma$-ray correlation and very rapid $\gamma$-ray variability.

Mrk 421 is a high frequency peaked BL Lac object (HBL) and at a redshift of z=0.03. It is one of the closest and brightest sources in the extragalactic TeV sky. It is the first extragalactic TeV source \citep{pun92} and has been targeted by many multi-wavelength campaigns. Mrk 421 is a highly variable source and show strong variability in both X-ray and $\gamma$-ray bands. The synchrotron peak of its SED ranges from 0.1keV in a low state to several keV in a high state. In general, the spectrum becomes harder with increasing flux level in both X-ray and $\gamma$-ray bands \citep{aha02,fos08,acc11}. In addition, the spectral index changes during different active states and shows an evident correlation with flux \citep{kre02}. Some observations have been revealed that Mrk 421 shows the hard and soft lag in the X-ray band \citep{tak96,tak00}.  The correlation of X-ray and $\gamma$-ray fluxes were found \citep{ino96,tak96,fos08,hor09},
supporting the X-ray and $\gamma$-ray emission may originate from the same electron population, while no apparent correlation of optical and X-ray appeared \citep{alb07,gie07,hor09,acc11}, indicating the optical and X-ray radiation may arise from the different electron population.
The rapid flares in both X-ray and $\gamma$-ray can help us to understand the particle acceleration and emission process in blazars.
First-order Fermi process may play an important role in accelerating the jet electrons \citep{mas97,kir98}. The electrons in the jet of Mrk 421 are likely to be accelerated by a first-order Fermi mechanism at the shock front and then form a power-law energy spectrum with the index of $p\simeq2.2$ \citep{abd11,shu12}. Several authors also pointed out that the stochastic acceleration could be at work in blazar jets \citep{lef11,yan12}. Recently, \citet{zhe11} suggested that the rapid TeV flares in Mrk 501 could be caused by the stochastic acceleration. It is likely that several acceleration mechanism are at work in blazar jets \citep{rie07}.

Motivated by above results, we study the flares of Mrk 421 on 2008 June 6-15, in which no significant correlation between optical and X-ray/TeV bands is observed. We assume that the multi-band emission originates from two different components. One is the steady component from the out region to produce the optical emission, in which the electrons are accelerated by first-order Fermi process. The other is the variable component from the inner region, in which the electrons are accelerated by stochastic process. The observed flares are produced by the hardening of the electron spectrum under the process of stochastic acceleration.
In the section 2 we describe the basic option of the model. In the section 3 we apply the model to Mrk 421. The discussion and conclusion are presented in the session 4. Throughout this paper, we adopt the cosmological parameters of $H_0 = 70 km s^{-1} Mpc^{-1}$, $\Omega_m = 0.3$, $\Omega_{\Lambda} = 0.7$.

\section{Model}
We use the conventional SSC model to produce the emission from the steady component. First-order Fermi acceleration can produce a paw-law electron energy spectrum. We assume the electron spectrum with the form
\begin{equation}
N'(\gamma')=k\gamma'^{-p} \mbox{$ \quad\quad \gamma'_{min}<\gamma'<\gamma'_{max}$},\\
\end{equation}
where $k$ is the normalization factor in the unit of $cm^{-3}$, $\gamma'_{min}$ and $\gamma'_{max}$ are the minimum and maximum energies of  electrons in the blob, and $p$ is the energy spectral index of electrons. Throughout the paper, the primed quantities refer to the co-moving frame and the unprimed quantities refer to the observed frame. Photon and electron energies are reduced in the unit of $m_ec^2$.

The synchrotron flux is given by \citep{fin08}
\begin{equation}
f_{\epsilon}^{syn}=\frac{\sqrt{3}\delta_{D}^{4}\epsilon'e^{3}B'}{4\pi hd_{L}^2}\int_{1}^{\infty}d\gamma'N'(\gamma')(4\pi{R'_b}^3/3)R(x),
\end{equation}
where $e$ is the electron charge, $B'$ is the magnetic strength, $h$ is the planck constant, $\delta_D$ is the Doppler factor, $R'_b$ is the radius of the emitting blob, and $d_{L}$ is the luminosity distance. Here, $\epsilon=\delta_{D}\epsilon'/(1+z)$ is the dimensionless synchrotron photon energy in the observer's frame. We use an approximation for $R(x)$ given by \citet{fin08}, where
$x=4\pi\epsilon'm_e^2c^3/3eBh\gamma'^2$. The synchrotron energy density is given by
\begin{equation}
u'_{syn}(\epsilon')=\frac{R'_b}{c}\frac{\sqrt{3}e^{3}B'}{h}\int_{1}^{\infty}d\gamma'N'(\gamma')R(x).
\end{equation}

The SSC flux is given by \citep{fin08,der09}
\begin{eqnarray}
f_{\epsilon_{s}}^{ssc}=\frac{9\sigma_{T}{\epsilon'_{s}}^2}{16\pi {R'_{b}}^2}\int_{0}^{\infty}d\epsilon'\frac{f_\epsilon^{syn}}{\epsilon'^{3}}
\int_{\gamma'_{min}}^{\gamma'_{max}}
d\gamma'
\nonumber\\
\frac{N'(\gamma')  (4\pi {R'_b}^2)/3  }{\gamma'^{2}}F(q,\Gamma_e),
\end{eqnarray}
where $\sigma_{T}$ is the Thomson cross section, and $\epsilon_s=\delta_{D}\epsilon_s'/(1+z)$ is the dimensionless scattered photon energy in the observer's frame. The function $F(q,\Gamma_e)$ is given by
$F(q,\Gamma_e)=[2qlnq+(1+2q)(1-q)+\frac{1}{2}\frac{(\Gamma_e q)^2}{(1+\Gamma_e q)}(1-q)]$ $(\frac{1}{4\gamma'}<q<1)$,
where $q=\frac{\epsilon'_s/\gamma'}{\Gamma_e(2-\epsilon'_s/\gamma')}$ and $\Gamma_e=4\epsilon'\gamma'$.

We use the time-dependent one-zone SSC model to produce the emission from the variable component under the process of stochastic acceleration. The momentum diffusion equation describing the evolution of particle number density $N'(\gamma',t)$ is given by \citep{kat06,tra11}
\begin{eqnarray}
\frac{\partial N^{\prime}(\gamma^{\prime},t)} {\partial t}  =
\frac{\partial}{\partial \gamma^{\prime}}
\left\{C^{\prime}(\gamma^{\prime},t)-A^{\prime}(\gamma^{\prime},t)\right\}N^{\prime}(\gamma^{\prime},t)
\nonumber\\
+D_p^{\prime}(\gamma^{\prime},t)\frac{\partial N^{\prime}(\gamma^{\prime},t)} {\partial \gamma^{\prime}}
-E^{\prime}(\gamma^{\prime},t) + Q^{\prime}(\gamma^{\prime},t),
\end{eqnarray}
where $C'(\gamma',t)$ is the radiative cooling term due to the synchrotron and SSC emission of the particle, $D'_p(\gamma',t)=\gamma'^2/2t'_{acc}$ is the momentum diffusion coefficient, $A'(\gamma',t)=2D'_p(\gamma',t)/\gamma'$ is the acceleration term which describes the particle energy gain per unit time, and $E'(\gamma',t)=N'(\gamma',t)/t'_{esc}$ is the escape term. The escape of the particles is described by the characteristic escaping time $t'_{esc}=R'_b/c$. $Q'(\gamma',t)$ is the particle injecting term.

In the framework of the quasi-linear approximation for the particle-wave interactions, the momentum diffusion coefficient is given by
\citep{sch89,sta08,osu09}
\begin{eqnarray}
D'_p\simeq\beta^2_A(\frac{\delta B'}{B'})^2(\frac{\rho'_g}{\lambda'_{max}})^{q-2}(\frac{\lambda'_{max}}{c})^{-1}p'^2,
\end{eqnarray}
where $\beta_A=v_A/c$ and $v_A$ is Alfv\'en wave velocity, $q$ is the spectral index of turbulence. $\rho'_g=p'c/eB=\gamma' mc^2/eB$ is the Larmor radius, $p'$ is the momentum of particle. $\lambda'_{max}$ is the maximum wavelength of the Alfv\'en wave spectrum. $\delta B'^2/B'^2$ describes the turbulence level. The characteristic acceleration timescale due to stochastic particle-wave interactions is
\begin{eqnarray}
t'_{acc}\simeq\frac{p^2}{D'_p}=(\frac{c}{v_A})^2(\frac{B'}{\delta B'})^2(\frac{\lambda'_{max}}{c})(\frac{\rho'_g}{\lambda'_{max}})^{2-q}\propto\gamma'^{2-q}.
\end{eqnarray}
In the case of hard sphere approximation ($q=2$), the acceleration timescale is independent with the electron energy.
We assume that the turbulence level $\delta B'^2/B'^2\sim1$, Alfv\'en wave velocity $v_A\sim c$ and the maximum wavelength $\lambda'_{max}\sim R'_b$. We can obtain $t'_{acc}\sim R'_b/c$ and assume $t'_{acc}=t'_{esc}=R'_b/c$ in our model. Here, we consider that the particle is continuously injected at the low energy($1\leq\gamma'\leq2$) and is accelerated to the equilibrium  energy $\gamma'_{eq}$ given by $t'_{cool}(\gamma'_{eq})=t'_{acc}$.

The total cooling rate is
\begin{equation}
C'(\gamma',t)=(\frac{d\gamma'}{dt})_{syn}+(\frac{d\gamma'}{dt})_{ssc}. \nonumber
\end{equation}

For the synchrotron cooling,
\begin{equation}
(\frac{d\gamma'}{dt})_{syn}=\frac{4\sigma_Tc}{3m_ec^2}\gamma'^2U'_B,\nonumber
\end{equation}
where $U'_B=\frac{B'^2}{8\pi}$ is the magnetic field energy density.

For the IC cooling, the Klein-Nishina (KN) effect is important at high energy and obviously modifies the electron and the SSC spectra  \citep{mod05}. We calculate the IC cooling rate using the method of \citet{mod05}, which takes into account the KN correction
\begin{equation}
(\frac{d\gamma'}{dt})_{SSC}=\frac{4\sigma_Tc}{3m_ec^2}\gamma'^2 \int f_{KN}(4\gamma'\epsilon')u'_{syn}(\epsilon')d\epsilon',\nonumber
\end{equation}
where $u'_{syn}(\epsilon')$ is given by the equation (3) and $f_{KN}(x)$ is approximated as \citep{mod05}
\begin{equation}
f_{KN}(x)\simeq\left\{
    \begin{array}{ll}
    (1+x)^{-1.5}    &\mbox{$For \quad x<10^4$}\\
     \frac{9}{2x^2}(\ln x- \frac{11}{6} ) &\mbox{$For \quad x>10^4$}\\
    \end{array}
\right.
\end{equation}

To obtain the electron energy distribution, Equation 5 needs to be solved by a numerical method because of its nonlinearity. We adopt a very useful numerical difference scheme proposed by \citet{cha70}. The method is a finite difference scheme, which uses the forward differentiation in time and the centered differentiation in energy. We test our code by comparing our results with the analytic solution given by \citet{cha70}. We find that our numerical results are in good agreement with the analytic solution.

\section{Application}

Mrk 421 entered a very active phase and showed frequent flare episodes in 2008. \citet{don09} presented the optical, X-ray, and very high energy $\gamma$-ray observation of Mrk 421 between May 24 and June 23 of 2008 with WEBT, UVOT, RXTE/ASM, Swift, AGILE, VERITAS and MAGIC. SuperAGILE, RXTE/ASM and Swift/BAT observe a clear correlation  between soft and hard X-rays. The $2-10$ keV flux measured by Swift/XRT is $\sim2.6\times10^{-9}erg\cdot cm^{-2} \cdot s^{-1}$, higher than previous observation. The derived peak synchrotron energy of $\sim3$ keV is higher than the typical values of $0.5-1$ keV . VHE observations with MAGIC and VERITAS between June 6 and 8 imply that $\gamma$-rays is well correlated with X-rays. Two flare states are reported during this period. One is the low state on June 6 observed from optical to TeV gamma-ray. The one is the high state on June 9-15 observed from optical to MeV gamma-ray, in which the flux increases and the spectrum becomes harder. Unfortunately, there are no TeV data included in this multi-wavelength compaign after June 8 because the moonlight hampers the Cherenkov telescope measurements. However, \citet{di10} complemented the VHE observation in the high state by the ARGO-YBJ experiment on June 11-13. We use the multiwavelenth data collected by \citet{don09} in June 6-15. The VHE data in the high state are from \citet{di10}. The TeV data is corrected by EBL absorption using the model reported by \citet{rau08}.

\begin{table*}
\caption{The model parameters for the steady component.} \label{para}
\begin{center}
\begin{tabular}{lccccccccc}
\hline

& $B'(G)$ & $\delta_D$ & $\gamma'_{min}$ & $\gamma'_{max}$ & $p$ & $k(cm^{-3})$  & $R'_b(cm)$\\

 \hline

The steady component & 0.2 & 15 & $1\times10^2$ & $4\times10^4$ & 2.2 &$9.2\times10^3$ &$1.5\times10^{16}$
\\\hline
\end{tabular}
\end{center}
\end{table*}

\begin{table*}
\caption{The model parameters for the variable component.}
\begin{center}
\begin{tabular}{lccccccccc}
\hline
& $B'(G)$ & $\delta_D$ & $Q'(cm^{-3}\cdot s^{-1})$ & $t'_{esc}(s)$ & $t/t'_{acc}$  & $R'_b(cm)$  \\
\hline
The variable component  \\
The low  state & 0.3 & 32 & $1.2\times10^{-3}$ & $R'_b/c$  &$5.74$ &$2.85\times10^{15}$    \\
The high state & 0.3 & 32 & $1.2\times10^{-3}$ & $R'_b/c$  &$6.50$ &$2.85\times10^{15}$    \\
\hline
\end{tabular}
\end{center}
\end{table*}

\begin{table*}
\caption{Jet powers in the form of radiation, Poynting flux, bulk motion of electrons and
 protons for the steady and variable components.}
\begin{center}
\begin{tabular}{lccccccccc}
\hline
& $P_r(erg\cdot s^{-1})$ & $P_B(erg\cdot s^{-1})$ & $P_e (erg\cdot s^{-1})$ & $P_p(erg\cdot s^{-1})$ & $\kappa_{eq}$  \\
\hline
The steady component    & $3.00\times10^{42}$ & $7.59\times10^{42}$ & $4.99\times10^{43}$ & $2.19\times10^{44}$  &$0.15$ \\
The variable component  \\
the low  state          & $3.70\times10^{42}$ & $2.81\times10^{42}$ & $1.61\times10^{43}$ & $1.37\times10^{44}$  &$0.17$ \\
the high  state         & $5.50\times10^{42}$ & $2.81\times10^{42}$ & $1.97\times10^{43}$ & $1.38\times10^{44}$  &$0.14$ \\
\hline
\end{tabular}
\end{center}
\end{table*}

\begin{figure*}
\begin{center}
\FigureFile(140mm,140mm){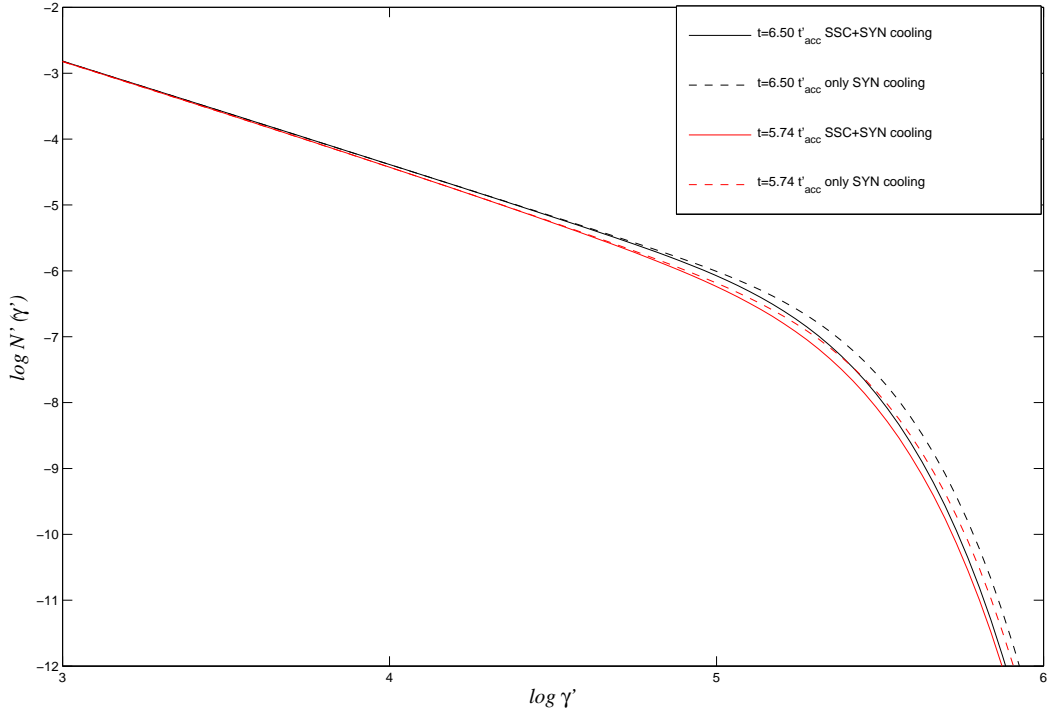}
\end{center}
\caption{ The electron spectra of two flare states used in the SED fittings. Solid lines represent the case of SSC cooling.
The dashed lines represent the case of only synchrotron cooling. \label{fig1} }
\end{figure*}

\begin{figure*}
\begin{center}
\FigureFile(140mm,140mm){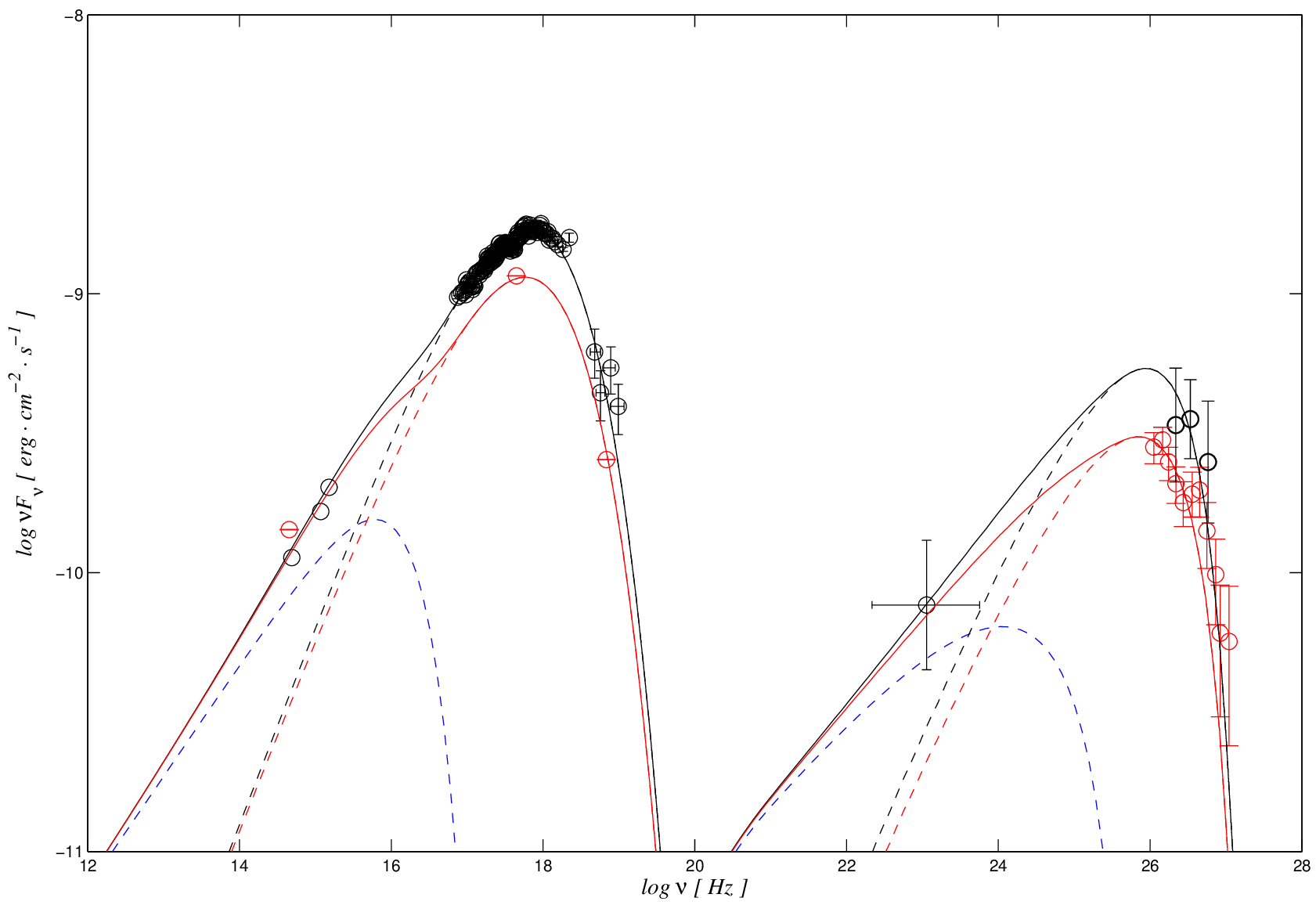}
\end{center}
\caption{ The red and black circles represent the simultaneous optical-X-ray-TeV data during two flare states. The blue dashed curves represent the synchrotron-SSC emission from the steady components. The red and black dashed curves represent synchrotron-SSC emission from the variable component during two flare states. The red and black solid curves are the total emission from two components during two flare states. The observed data come from \cite {don09} and \cite {di10}.  \label{fig2} }
\end{figure*}

Many multiwavelenth campaigns have revealed a strong correlation of  X-rays with $\gamma$-rays.  No significant correlation is observed between X/$\gamma$-rays and the optical band which usually shows a steady flux. We argue that the emission may contain two components from two different jet regions. One is the steady component from the outer region, in which the electrons are accelerated by first-order Fermi process. We use the conventional SSC model to produce the emission from the steady component. The first-order Fermi acceleration is expected to create a steady electron spectrum with spectral index 2.2, we set $p=2.2$ to fit the SED of Mrk 421, in which the parameters are listed in table 1. The other is the variable component from the inner region, in which the low energy electrons are continuously injected into blob and then accelerated to the high energy by the stochastic acceleration. In our calculations, We assume a constant initial electron distribution $N'_{ini}(\gamma',0)=1.2\times10^{-3}cm^{-3}$ for $1\leq\gamma'\leq2$. We use the time-dependent SSC model to produce the emission from the variable component, in which the parameters are listed in table 2. The model SED of Mrk 421 is shown in figure 2. It shows that our model can well fit the SEDs of Mrk 421.

In figure 1, we show the electron spectra of two flare states used in our SED fits. It is obvious that the IC cooling modifies the electron spectrum in the high energy and the electron spectrum is softer than that in the case of only synchrotron cooling. Nevertheless, the electron spectrum does not change at the low-energy end and becomes harder at the high-energy end from the low state to the high state, indicating that the flares are due to the hardening of the electron spectrum at the high energy. The short time variability ($\sim1$ day) constrains the size of the emitting region to be $R'_b<ct_{var}\delta_D\sim5\times10^{16}(\delta_D/20)cm$. The mean energy $E$ of synchrotron radiation from an electron with Lorentz factor $\gamma'$ is $E\simeq\frac{2eB'\gamma'^2\delta_D}{3\pi h m_ec}$. The peak energy of synchrotron radiation is $E_{s}\simeq3$ keV during this flare, corresponding to $\gamma'_{b}\simeq 1.4\times10^5(B/0.3G)^{-1/2}(\delta_D/32)^{-1/2}$, which is comparable with the peak of the electron spectrum we used (see fig.1). The parameter of $\gamma'_b\epsilon'_s=\gamma'_b\epsilon_s/\delta_D\gg1$ indicates that the IC scattering occurs at the KN regime during the two flares. We can estimate the Doppler factor $\delta_D$ by the KN effect. When the energy of the electron has $\gamma'\geq1/\epsilon'_s$, the IC scattering takes place at the KN regime, leading to a peak in the IC luminosity at $\epsilon'_c\simeq1/\epsilon'_s$. We can obtain that
$\delta_D\simeq(\epsilon_s\epsilon_c)^{\frac{1}{2}}=(\frac{E_s}{m_ec^2}\frac{E_c}{m_ec^2})^{\frac{1}{2}}$. We have $E_s\simeq1$ keV  and $E_c\simeq0.3 $ TeV in the low state and obtain $\delta_D\simeq34$ that is very close to the value we used. The physical parameters we used in the inner region are very similar to that of \citet{gie07}. It is indicated that a high Doppler factor is required to reproduce the observed SED \citep{gie07,fin08,ale12}.

We can estimate the power carried by the jet in the form of Poynting flux ($P_B$), relativistic electrons ($P_e$) and cold protons ($P_p$). All powers are calculated as \citep{cel08,ghi10}
\begin{equation}
P_i=\pi R'{_b^2}\Gamma^2cU'_i,
\end{equation}
where $U'_i$ $(i=B,e,p)$ is the energy density of the i component as measured in the co-moving frame. The electron energy density is given by
$U'_e=\int\gamma' m_ec^2N'(\gamma')d\gamma$. The ratio of electron number density to proton number density in blazar jets is of the order of 0.1-1  \citep{sik00,cel08}. We calculate the proton energy density $U'_p$ by assuming one proton per emitting electron.
The power carried by the radiation is \citep{cel08}
\begin{equation}
P_r=\pi R'{_b^2}\Gamma^2cU'_r=L\frac{\Gamma^2}{\delta^4}\simeq\frac{L}{\delta^2},
\end{equation}
where $U'_r=L'/(4\pi R'{_b^2}c)$ is the radiation energy density produced by the jet. $L$ is the total observed non-thermal luminosity. We set $\delta=\Gamma$ by assuming the viewing angle $\theta=1/\Gamma$. $\kappa_{eq}=U'_B/U'_e$ is the equipartition parameter between the magnetic and electron energy density. The estimated powers are listed in table 3. in both regions, we find that the electron energy density is slightly larger than the magnetic field energy density, and they are  close to equipartition, typically $0.1\lesssim\kappa_{eq}\lesssim1$ \citep{ghi10}. Because the radiated power is far less than the total jet power, the substantial fraction of the total jet power is carried by the proton, indicating that the jet is strongly matter-dominated.

\section{Discussions and Conclusions}
\citet{don09} analyzed the optical and X-ray variability of Mrk 421, they suggested that the inner region would produce the X-ray. However outer region only could produce lower-frequency emission. \citet{bla05} pointed out that the one-zone SSC model could't fit the optical and radio fluxes well. They argued that the optical and radio radiation could originate in another region further down the jet, which is consistent with the fact that the optical flux has small change compared to the fluxes in the X-ray and $\gamma$-ray bands. Moreover, a two-component model is used to explain the multiwavelength variability of Mrk 421 \citep{che11}. A two-component model is seemly needed to explain the multi-wavelength SEDs of Mrk 421, supported by the lack of correlation between the optical and X-ray bands.

In the paper, we study the X-ray and $\gamma$-ray flares of Mrk 421 on 2008 June 6-15 using the SSC model with electron acceleration. We suggest that the low-energy electron population from the outer region is produced via first-order Fermi acceleration, and the high-energy electron population is related to the stochastic acceleration.
The X-ray observations with Suzaku also implied that the emission from Mrk 421 may originate from two different electron populations, which may be produced by fist-order Fermi and second-order Fermi processes \citep{ush09,ush10}.
We reproduce the  observed SEDs of Mrk 421 from the low state to the high state using a two-component model. We obtain an excellent fit to the observed spectrum of Mrk 421 based on our model.
\citet{don09} used a steady electron spectrum to reproduce the observed SEDs of Mrk 421, and pointed out that the flares are due to the softing/hardening of electron spectrum, not due to the increase/decrease of the electron density.
However, they did't explain what acceleration mechanism leads to the softing/hardening of electron spectrum. We argue that the flares are due to the hardening of electron spectrum at the high energy under the process of stochastic acceleration. In our model, the electrons with low energy are assumed to be continuously injected into the blob and are then accelerated to the higher energy by stochastic acceleration, which leads to the hardening of the electron spectrum and then causes the X-ray and $\gamma$-ray flares by the SSC process. \citet{gar10} also suggested that the X-ray flares of Mrk 421 are due to the intrinsic changes in the acceleration process, which is consistent with our results.  Furthermore, the observed  peaks of synchrotron and IC components  shift toward the higher frequency from the low state to the high state, supporting that the flares are driven by the process of the electron acceleration. We  estimate the jet powers using our model parameters in both jet regions. We find that the energy densities of electrons and magnetic field are near equipartition and the jet is strongly matter-dominated in both jet regions. A hint of correlation between optical and TeV energy was reported by \citet{don09}, but no clear correlation between the optical and X/$\gamma$-ray bands is confirmed so far. Our two-component model can be tested by observing the optical and X/$\gamma$-ray bands in future.
\\

We thank the anonymous referee for valuable comments and suggestions.
We acknowledge the financial supports from the National Basic Research Program of China
(973 Program 2009CB824800), the National Natural Science Foundation
of China 11133006, 11163006, 11173054, and the Policy Research Program of Chinese Academy of
Sciences (KJCX2-YW-T24).

%%%
% See the manual for the detail.
%%%

%%%%%%%%%%%%%%%%%%%%%%%%%%%%%%%%%%%%%%%

\end{document}